\documentclass[twocolumn,showpacs,preprintnumbers,amsmath,amssymb]{revtex4}

\usepackage{graphicx} 
\usepackage{dcolumn} 
\usepackage{bm} 

\begin{document}

\title{Pb chains on reconstructed Si(335) surface}

\author{Mariusz Krawiec} 
  \email{krawiec@kft.umcs.lublin.pl}
\affiliation{Institute of Physics, M. Curie-Sk\l odowska University, 
             Pl. M. Curie-Sk\l odowskiej 1, 20-031 Lublin, Poland}

\date{\today}

\begin{abstract}
The structural and electronic properties of Si(335)-Au surface decorated with 
Pb atoms are studied by means of density-functional theory. The resulting
structural model features Pb atoms bonded to neighboring Si and Au surface 
atoms, forming monoatomic chain located 0.2 nm above the surface. The presence
of Pb chain leads to a strong rebonding of Si atoms at the step edge. The fact
that Pb atoms occupy positions in the middle of terrace is consistent with STM
data, and also confirmed by simulated STM images. The calculated band structure 
clearly shows one-dimensional metallic character. The calculated 
electronic bands remain in very good agreement with photoemission data.
\end{abstract}

\maketitle


\section{\label{intro} Introduction}


One-dimensional (1D) atomic chains created on semiconductor templates have 
recently attracted much attention due to new phenomena characteristic for a 
reduced dimensionality \cite{Giamarchi}. The most spectacular examples include 
a breakdown of Fermi liquid theory \cite{Luttinger,Auslaender} and Peierls
metal-insulator transition \cite{Peierls}. A specially attractive route to
create such 1D structures is a process of self-organization of atoms into very
regular arrays of parallel metallic chains on stepped semiconducting or
insulating substrates \cite{Himpsel,Owen}. In this case electrons near the 
Fermi energy are completely decoupled from the substrate due to band gap in 
electronic spectrum of the substrate. The chain atoms are bounded to the 
surface by low energy states which do not contribute to electronic properties 
of the system.

The Si(335)-Au surface is one of the simplest examples of high-index surfaces
which stabilize the one-dimensional structures, and has been studied by number 
of techniques, including: reflection high energy electron diffraction (RHEED) 
\cite{Zdyb}, scanning tunneling microscopy (STM) \cite{Crain,MK_1}, 
angle-resolved photoemission spectroscopy (ARPES) \cite{Crain,Kisiel} and first 
principles density functional theory (DFT) \cite{MK_2}. In particular, the STM 
topography data show regular arrays of monoatomic chains separated by width of 
Si(335) terrace and a few nanometers long \cite{Crain,MK_1}. The ARPES spectra
taken in the direction parallel to the steps show two highly dispersive bands 
crossing the Fermi energy ($E_F$), thus indicating one-dimensional metallic 
nature of the system \cite{Crain}. The structural model of the Si(335)-Au 
surface has also been proposed \cite{Crain} and confirmed later by the DFT
calculations \cite{MK_2}, which well describes all the available experimental 
data.

Recently, other materials (Na, Pb, In) deposited on Au decorated Si(335) 
surface have also been studied 
\cite{Starowicz,Jalochowski,Kisiel_2,Zdyb_2,Skrobas,Zdyb_3}. Perhaps lead is 
the most intensively studied among them. Unlike Si(335)-Au surface, in which 
gold substitutes Si atoms in the surface layer, the Si(335)-Au/Pb is a 
representative of new class of systems in which the deposited material forms 
one-dimensional structures adsorbed on-top of the surface. The lead deposited 
on flat Si(111)-Au(6$\times$6) surface forms well ordered monoatomic layers 
\cite{Jalochowski_2,Jalochowski_3,Jalochowski_4}, and is known to weakly 
interact with the Si substrate \cite{Dil}. Thus, in the case of Si(335)-Au/Pb 
structure, one can expect a more pronounced one-dimensional character of chains 
as compared to clean Si(335)-Au reconstruction. Indeed, the deposition of 
0.28 ML of Pb on reconstructed Si(335) surface (0.28 ML gives exactly 1 Pb atom 
per Si(335) unit cell) also leads to one-dimensional objects on the surface. 
The STM topography shows a few nanometers long chains placed in between Si 
chains of original Si(335)-Au surface \cite{Kisiel_2}. The photoemission 
spectra taken in the $[1 \bar{1} 0]$ direction (i.e. parallel to the steps), 
show a highly dispersive band crossing the Fermi energy and quite flat bands in 
the $[1 1 \bar{2}]$ direction (perpendicular to the steps), thus indicating 
clear one-dimensional character of the reconstruction. A more detailed analysis 
of morphology of the surface and of the ARPES data is difficult or even 
impossible without appealing to DFT calculations. It is the purpose of the 
present work to propose a structural model of the Si(335)-Au/Pb surface and 
calculate corresponding band structure.

Here I present a structural model of the Pb chains on Si(335)-Au surface,
derived from total energy DFT calculations. It features single Pb atom per
Si(335) unit cell placed near the Au chain. The Pb atoms are bonded to
neighboring Si and Au atoms, forming monoatomic chain located $\sim 0.2$ nm
above the surface. This picture is consistent with the STM topography data of 
Ref. \cite{Kisiel_2}, and also confirmed by simulated STM topography images. On 
the other hand, the calculated band structure for the present model clearly 
shows one-dimensional character, i.e. a strong dispersion in the direction 
parallel to the steps and their lack in the direction perpendicular to them, 
and remains in good agreement with the ARPES spectra of Ref. \cite{Kisiel_2}. 
The rest of the paper is organized as follows. In Sec. \ref{details} the 
details of calculations are presented. The structural and electronic properties 
of the clean Si(335)-Au surface are briefly discussed in Sec. \ref{clean}. The 
structural model of Pb chains, simulated STM topography images and the band 
structure are presented and discussed in Sec. \ref{model}, \ref{STM}, 
\ref{band}, respectively. The influence of step-edge buckling on electronic 
properties is discussed in Sec. \ref{buckling}. Finally, Sec. 
\ref{conclusions} contains some conclusions.


\section{\label{details} Details of calculations}


The calculations have been performed using standard pseudopotential density
functional theory and linear combination of numerical atomic orbitals as a 
basis set, as implemented in the SIESTA code 
\cite{Ordejon,Portal,Artacho,Soler,Artacho_2}. The local density approximation
(LDA) to DFT \cite{Perdew}, and Troullier-Martins norm-conserving
pseudopotentials \cite{Troullier} have been used. In the case of Pb and Au 
pseudopotentials, the semicore $5d$ states were included. A double-$\zeta$ 
polarized (DZP) basis set was used for all the atomic species 
\cite{Portal,Artacho}. The DZP utilizes two radial functions for each angular 
momentum and additional polarization shell. The radii of the orbitals for 
different species were following (in Bohrs): Si - 5.13 ($3s$), 6.59 ($3p$) and 
5.96 ($3d$), Au - 4.39 ($5d$), 6.24 ($6s$) and 5.79 ($6p$), Pb - 3.56 ($5d$), 
4.68 ($5f$), 5.30 ($6s$) and 6.48 ($6p$), and H - 5.08 ($1s$) and 4.48 ($2p$). 
A Brillouin zone sampling of 24 nonequivalent $k$ points, and a real-space grid 
equivalent to a plane-wave cutoff 100 Ry (up to 82 $k$ points and 300 Ry in the 
convergence tests) have been employed. This guarantees the convergence of the 
total energy within $\sim 2$ meV per atom in the supercell.

The Si(335)-Au/Pb system has been modeled by four silicon double layers and a
vacuum region of 18 \AA. All the atomic positions were relaxed except the 
bottom layer. The Si atoms in the bottom layer were fixed at their bulk ideal
positions and saturated with hydrogen. To avoid artificial stresses, the 
lattice constant of Si was fixed at the calculated value, 5.41 \AA. The atomic 
positions were relaxed until the maximum force in any direction was less than
0.01 eV/\AA.


\section{\label{clean} Clean Si (335)-Au surface}

\begin{figure}[h]
 \resizebox{0.9\linewidth}{!}{
  \includegraphics{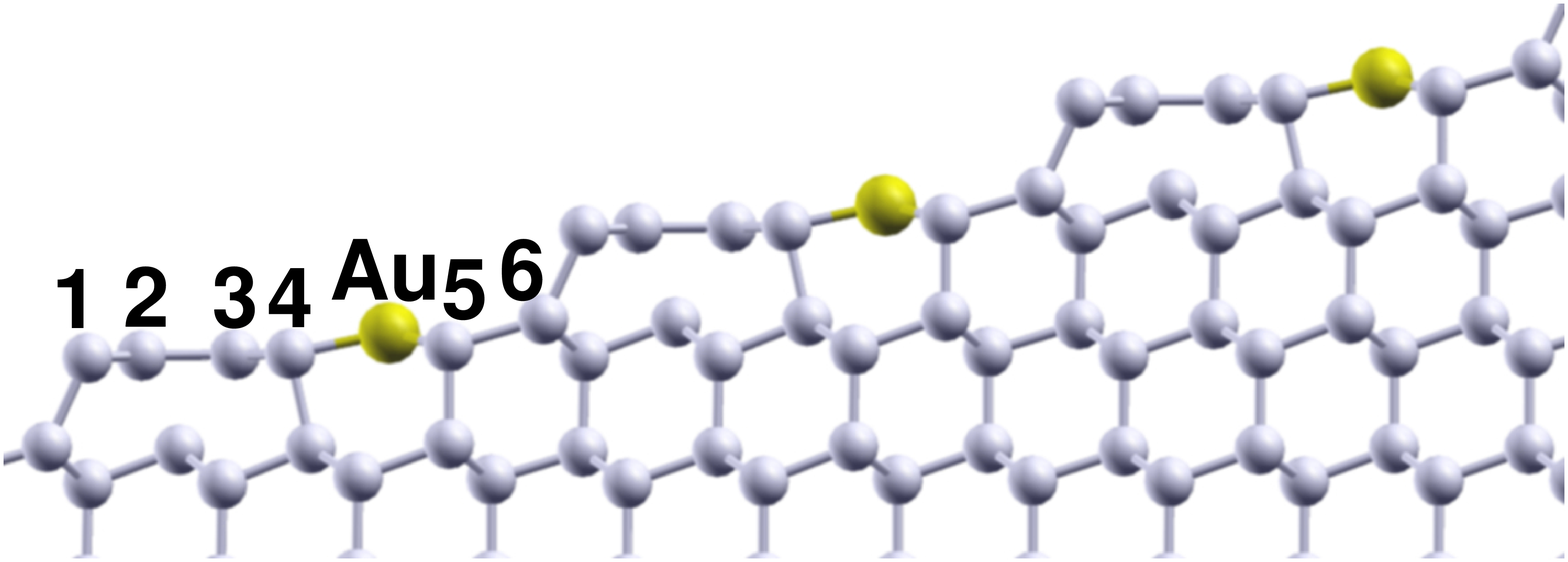}
} \\
 \vspace{0.5cm}
 \resizebox{0.9\linewidth}{!}{
  \includegraphics{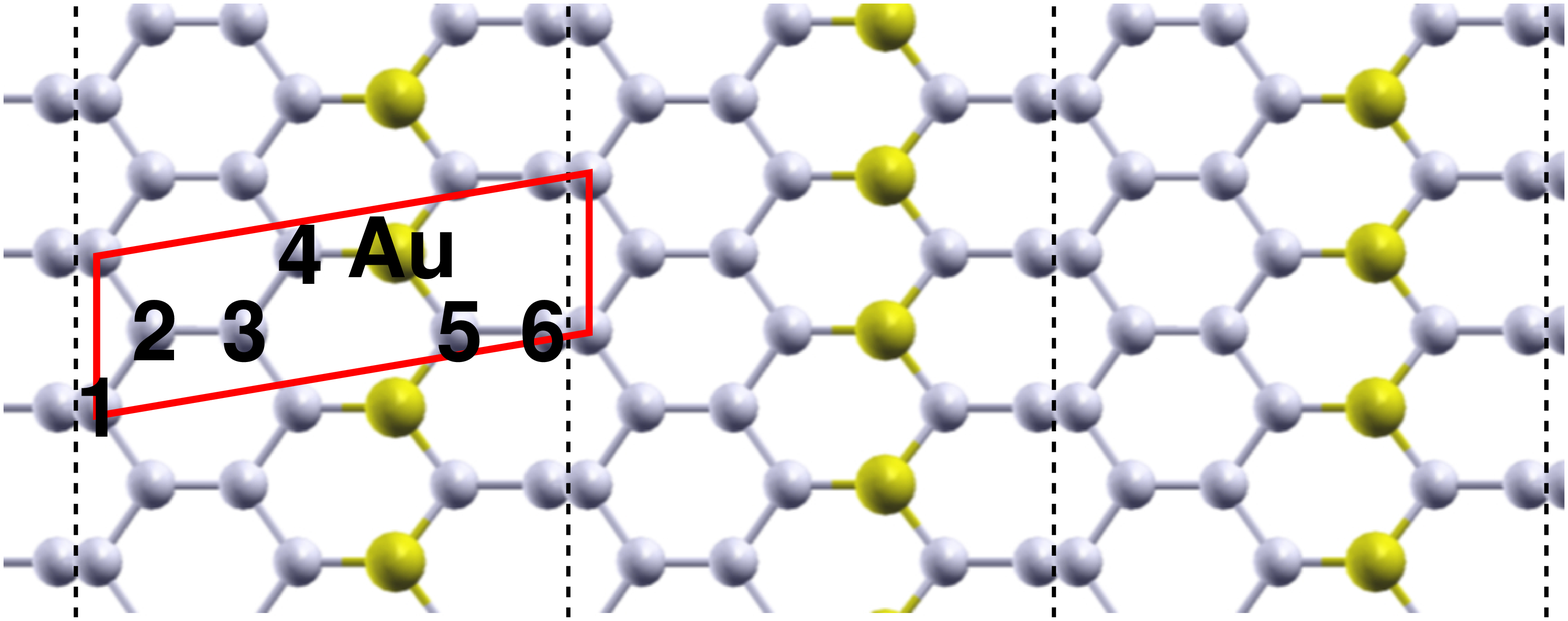}
}
 \caption{\label{Fig1} (Color online) Structural model of Si(335)-Au surface.
          Top panel shows side view of the structure, and bottom panel shows 
	  top view with marked surface unit cell. Labels 1-6 stand for silicon 
	  surface atoms (Si$_1$-Si$_6$), while Au denotes gold atom. The dashed 
	  lines in bottom panel indicate step edges.}
\end{figure}

The deposition of 0.28 ML of gold on Si(335) surface forms well ordered arrays 
of chain structure. The surface consists of (111) terraces which have a width 
$3 \frac{2}{3} \times a_{[1 1 \bar{2}]}$ (1.26 nm) \cite{Zdyb}. Each terrace 
contains a single row of gold atoms running parallel to the step edge, i.e. in
the $[1 \bar{1} 0]$ direction. The gold chain is formed by substitution of Si 
atoms in the middle of terrace. The step edge Si atoms form a 'honeycomb' 
substructure \cite{Erwin,Kang,Battaglia}, which is a common feature of all the 
Au-induced vicinal Si surfaces \cite{Crain}. The structural model of the 
Si(335)-Au surface is shown in Fig. \ref{Fig1}, where the Si surface atoms 
(Si$_1$-Si$_6$) are labeled by numbers 1-6, and the gold atom by Au.
Originally this model has been proposed as a simple truncation of Si(557)-Au
reconstruction \cite{Crain}, and later confirmed by DFT calculations
\cite{MK_2}. 

The STM topography data \cite{Crain,MK_1} show a single chain per terrace, 
which is associated with the step edge Si atoms rather than with the Au chain. 
The gold substitutes some of top layer Si atoms in the middle of terrace, and 
is not visible to STM. This has also been confirmed by DFT calculations 
\cite{MK_2}. See also simulated STM topography images presented in Fig. 
\ref{Fig3}. The ARPES spectra show two highly dispersive bands crossing the 
Fermi energy in the direction parallel to the steps, and quite flat bands in 
the direction perpendicular to them \cite{Crain}, clearly indicating 
one-dimensional metallic nature of the system. The DFT calculations indicate 
that one of the bands crossing $E_F$ is associated with the step edge Si atoms, 
having unsaturated dangling bonds, while the other one originates from 
hybridization of the Au and neighboring Si atoms in the middle of terrace in 
the surface layer \cite{MK_2}.


\section{\label{model} Structural model of Pb chains on Si(335)-Au}

\begin{figure}[h]
 \resizebox{0.9\linewidth}{!}{
  \includegraphics{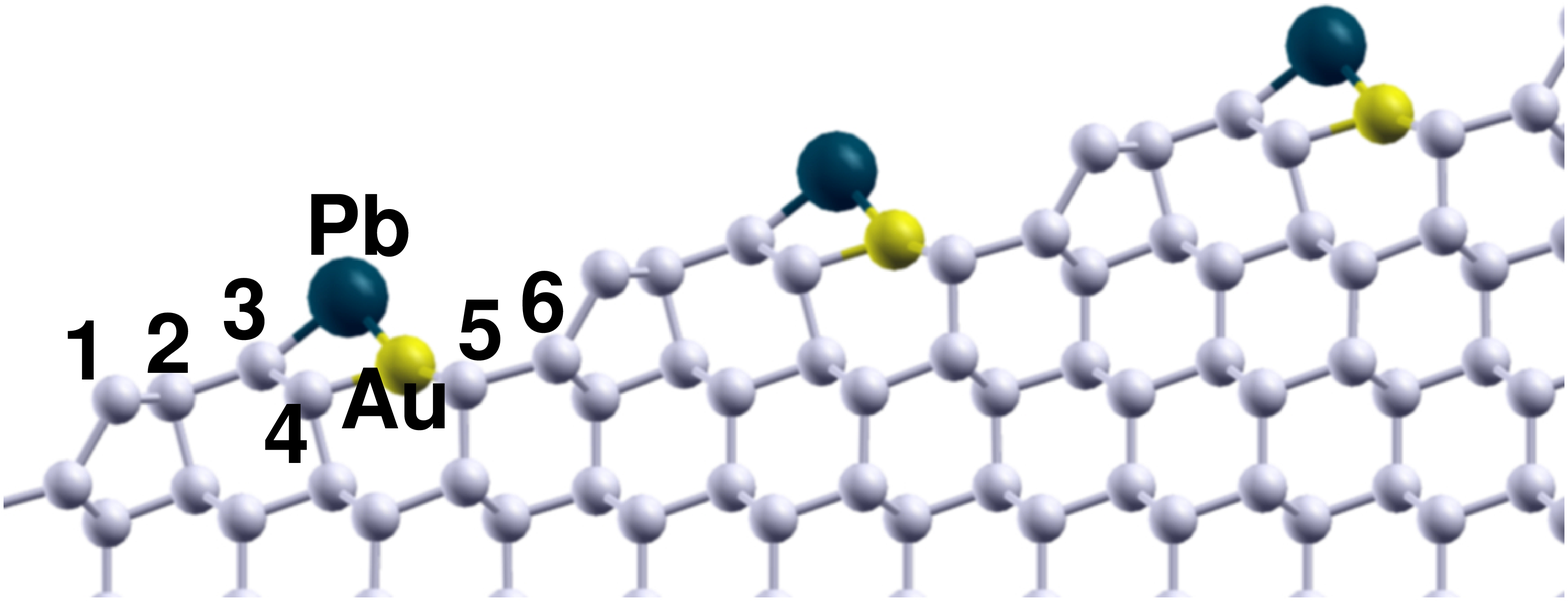}
} \\
 \vspace{0.5cm}
 \resizebox{0.9\linewidth}{!}{
  \includegraphics{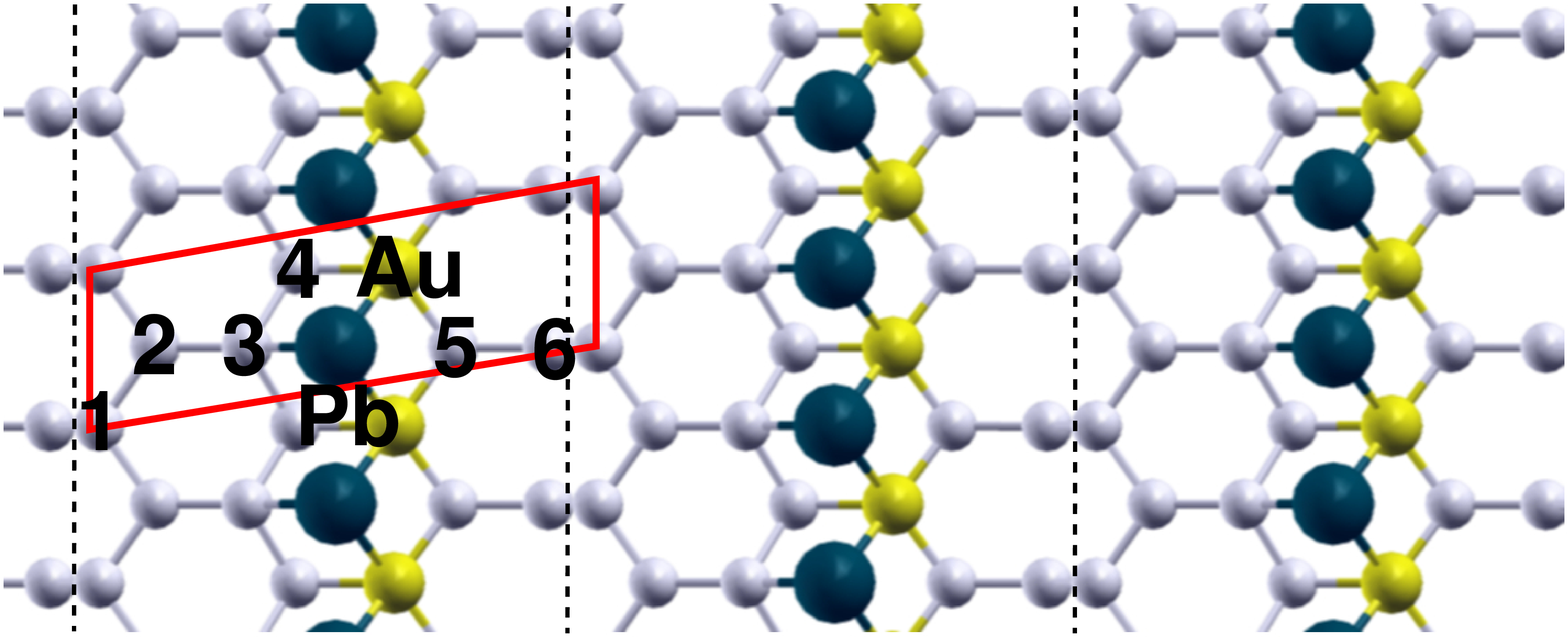}
}
 \caption{\label{Fig2} (Color online) Structural model of Si(335)-Au/Pb 
          surface. Top panel - side view, and bottom panel - top view of the 
	  model. Labels 1-6 stand for silicon surface atoms (Si$_1$-Si$_6$), 
	  while Au denotes gold and Pb - lead atom. Surface unit cell is also 
	  indicated in the bottom panel.
 }
\end{figure}

One-dimensional structures of Pb on Si(335)-Au reconstruction one obtains
assuming single Pb atom per Si(335)-Au unit cell, which corresponds to the 
experimental Pb coverage - 0.28 ML \cite{Kisiel_2}. The total energy 
calculations show that Pb atoms prefer to adsorb on the surface. The 
substitution of Pb atoms into top Si layer is energetically less favorable, as 
the surface energy is by 0.3-0.6 eV (per unit cell) higher in comparison to 
clean Si(335)-Au surface and Pb atom in the bulk fcc structure. Similarly, the 
substitution into the second Si layer is not preferred. In this case the energy 
cost is 1 eV. The exception is the substitution of Pb at the step edge, where 
the energy gain is 0.31 eV. More than 40 structural models have been 
investigated, and only five of them lead to stable structures. Corresponding 
surface energies, with respect to clean Si(335)-Au surface and bulk Pb atom, 
are shown in Table \ref{Tab1}. 
\begin{table}
\caption{\label{Tab1} The relative surface energies of most stable structural 
models of Si(335)-Au/Pb structure. The energies are referred to clean 
Si(335)-Au surface and Pb atom in the bulk fcc structure.}
\begin{center}
\begin{tabular}{ccccc}
\hline
& \vline & & \vline & \\
model & \vline & position of Pb & \vline & surface energy (eV)\\
& \vline & & \vline & \\
\hline
& \vline & & \vline & \\
1 & \vline & Si$_3$-Au-Au & \vline & $-0.41$ \\
2 & \vline & Si$_1$ (subst) & \vline & $-0.31$ \\
3 & \vline & Si$_5$-Si$_6$-Si$_1$ & \vline & $-0.25$ \\
4 & \vline & Au-Si$_1$-Si$_1$ & \vline & $-0.06$ \\
5 & \vline & Si$_4$-Au & \vline & $-0.02$ \\
& \vline & & \vline & \\
\hline
\end{tabular}
\end{center}
\end{table}
The differences in energy are rather small, however, as it will be argued 
later, the model with the lowest energy, i.e. Si$_3$-Au-Au model, is the best 
candidate for a true model of Pb chains on Si(335)-Au surface. Note the 
nomenclature used here reflects the bonding of lead with corresponding surface 
atoms. For labeling see Fig. \ref{Fig2}. The structural model with the lowest 
surface energy features Pb atoms located 2 \AA $\;$ above the surface near the 
gold chain, and is shown in Fig. \ref{Fig2}.  
As one can read off from the figure, lead is bounded to two Au atoms and one 
Si$_3$ atom, but not to Si$_4$ atom (Fig. \ref{Fig2}). The Pb-Si$_3$ bond 
length is equal 2.75 \AA, and Pb-Au - 2.92 \AA. Both bonds are shorter than the 
Pb bonds in the bulk fcc structure (3.46 \AA). The presence of Pb atoms on-top 
of the surface also modifies positions of underneath Au and Si atoms, leading 
to change in their bond lengths. Thus the Au-Si$_4$ bond is equal 2.52 \AA, and 
Au-Si$_5$ - 2.47 \AA, to compare with 2.43 \AA $\;$ and 2.38 \AA, respectively, 
in the clean Si(335)-Au surface. What is more important, presence of Pb atoms 
also leads to a strong rebonding at the step edge. In the clean Si(335)-Au 
surface, the Si atoms near the step edge form a sort of honeycomb chain (see 
Fig. \ref{Fig1}), which is common feature of all the vicinal Si surfaces 
\cite{Crain}. Main feature of this substructure is a true double bond between 
Si$_2$ and Si$_3$ atoms, which is responsible for the stability of the 
honeycomb chain \cite{Erwin}. In the presence of Pb atoms on Si(335)-Au 
surface, there is no longer double bonds between Si$_2$ and Si$_3$ atoms. The 
atoms are just single bonded, as in common silicon structures. The Si$_3$ atom 
is now bonded with the lead, and Si$_2$ with underneath Si atom in the second 
layer (compare Fig. \ref{Fig1} and Fig. \ref{Fig2}). 

At this point I would like to comment on main structural features of the other 
models, listed in Table \ref{Tab1}, which have slightly higher relative 
surface energies. The next 'best' structural model features the step edge Si
atoms substituted by lead (model 2 in Table \ref{Tab1}). Model 3, having 
relative surface energy -0.25 eV, accounts for Pb atom bonded to Si$_5$ and 
Si$_6$ atoms within the same terrace and Si$_1$ atom at the step edge of the 
neighboring terrace (see Fig. \ref{Fig2} for labeling). In model 4, with the 
energy -0.06 eV, Pb is located between two Si$_1$ atoms at the step edge and 
one gold atom on neighboring terrace, while in model 5 the Pb takes bridge 
position between Si$_4$ and Au atoms on the same terrace. However those models 
can be ruled out, according to arguments given below.

The present model, i.e. the model with the lowest relative surface energy, is a 
good candidate for structural model of Si(335)-Au/Pb reconstruction. There are 
few strong arguments supporting this model. First one is that the model has the 
lowest energy. The second one is that the model explains the STM data of Ref. 
\cite{Kisiel_2}. In STM images of Si(335)-Au/Pb surface, the Pb chains are 
located in between the chain structure of original Si(335)-Au reconstruction 
(see Fig. 1 of Ref. \cite{Kisiel_2}). According to the STM data, the Pb chains 
should be located in the middle of terraces, above the surface. This is exactly 
what the present model shows, and STM simulations, which will be presented in 
the next section, confirm this. The other models, listed in Table \ref{Tab1} 
remain in disagreement with the STM data. Model 1 leads to a single chain in 
STM topography, as the step edge Si atoms, which formed the monoatomic chains 
in clean Si(335)-Au surface are now replaced by the lead atoms. Similarly, 
models 2, 3 and 4 also lead to single chain within terrace, as the Pb atoms 
saturate the step edge Si dangling bonds. In fact, the STM topography of model 
4 shows a zig-zag chain, associated with Pb atoms and less visible step edge Si 
atoms. Only the last model, i.e. model 5, remains in reasonably good agreement 
with STM topography data. The next argument concerns the band structure, 
as the only present model reproduces the photoemission spectra of Ref. 
\cite{Kisiel_2} very well. The other models, which have slightly higher 
relative surface energies disagree with the ARPES data, in particular, they do 
not give correct band structure near the Fermi energy. This will be further 
discussed in Sec. \ref{band}. The next argument, albeit intuitive, is based on 
the analogy with the growth of Pb on flat Si(111) surface. It is well known, 
that lead grows in very regular fashion, i.e. layer by layer, on 
Si(111)-Au(6$\times$6) surface from very beginning 
\cite{Jalochowski_2,Jalochowski_3,Jalochowski_4}, contrary to the growth on 
Si(111) with 7$\times$7 reconstruction, where it forms amorphous wetting layer 
\cite{Jalochowski_2,Weitering,Budde}. This indicates that Pb prefers to bond 
with Au rather than with Si atoms. The present model of Si(335)-Au/Pb surface 
also reflects this fact, as the Pb atom is bonded with one Si and two Au atoms. 
The last argument comes from the conditions at which the experiment of Ref. 
\cite{Kisiel_2} has been performed. Namely, the temperature of deposition of Pb 
on Si(335)-Au surface was 260 K. This temperature is far to low to substitute 
the Si atom by the lead, and the model with Si$_1$ atoms replaced by Pb can be 
ruled out. All the above arguments show that the model shown in Fig. \ref{Fig2} 
is very good candidate to be a true model of the Pb chains on Si(335)-Au 
surface.


\section{\label{STM} STM simulations}

\begin{figure}[h]
 \resizebox{0.45\linewidth}{!}{
  \includegraphics{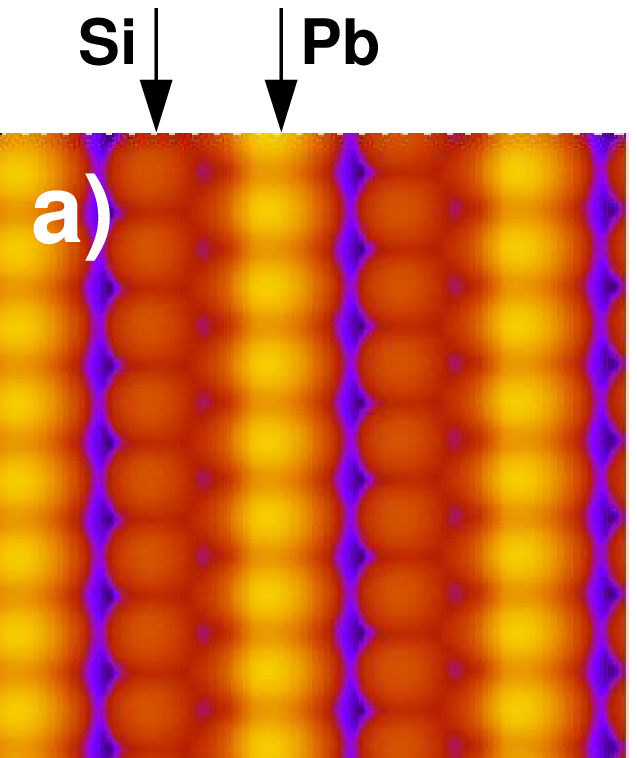}
} 
 \resizebox{0.45\linewidth}{!}{
  \includegraphics{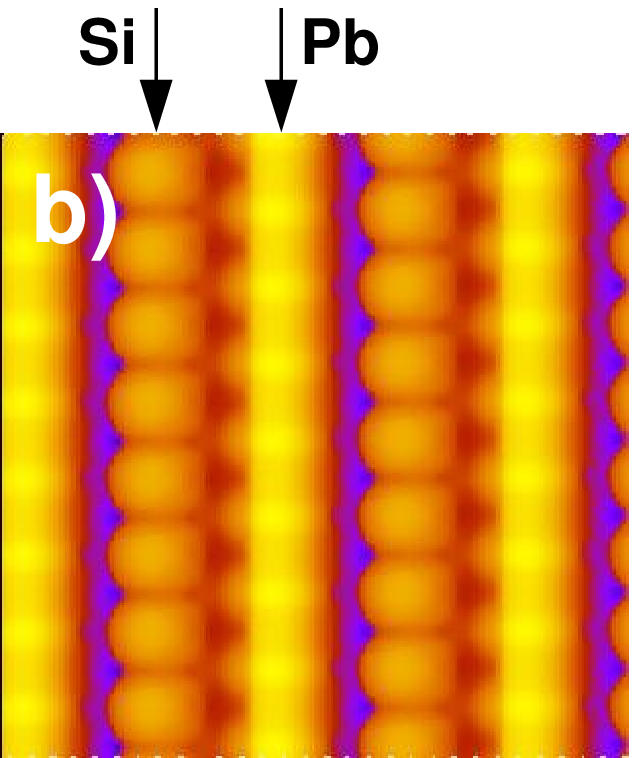}
} \\
 \vspace{0.05cm}
 \resizebox{0.45\linewidth}{!}{
  \includegraphics{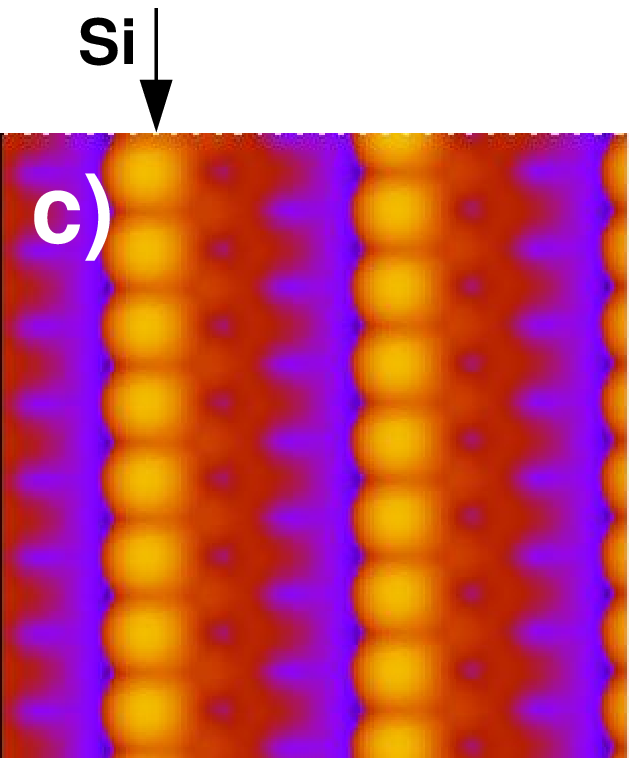}
}
 \resizebox{0.45\linewidth}{!}{
  \includegraphics{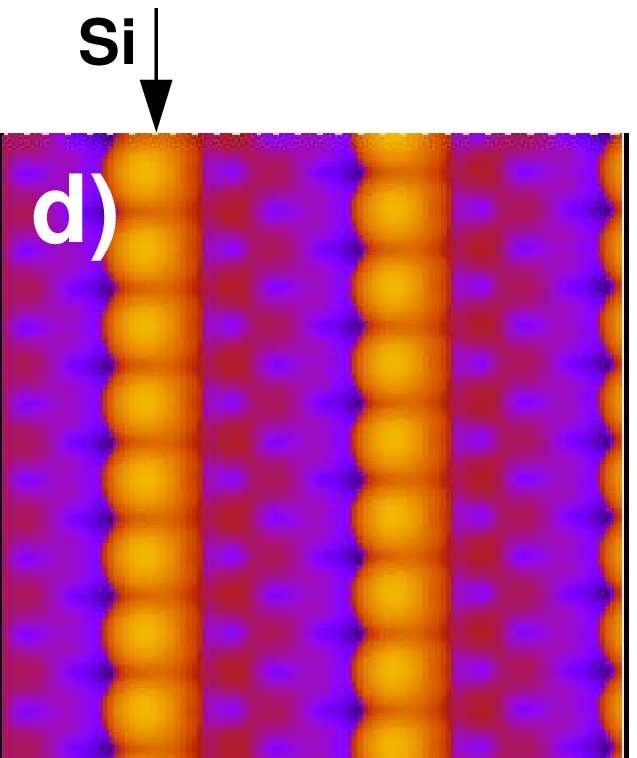}
}
 \caption{\label{Fig3} (Color online) STM simulations of 3 $\times$ 3 nm$^2$ 
          area of Si(335)-Au/Pb surface (top panels) and Si(335)-Au 
	  (bottom panels) for sample bias $U=-1$ V (a and c) and $U=-0.1$ V (b 
	  and d). All the images show the same area of the surface.
 }
\end{figure}

The STM topography data of clean Si(335)-Au surface shows one-dimensional
structures which are interpreted as the step edge Si atoms \cite{Crain,MK_2}.
The deposition of 0.28 ML of Pb leads to monoatomic Pb chains located in the 
middle of the Si(335)-Au terraces, i.e. between the Si chains of original 
Si(335)-Au surface. The structural model discussed in previous section supports
this scenario. To further check the validity of the structural model, I have
performed STM simulations within the Tersoff-Hamann approach \cite{Tersoff}.
The results of constant current topography for different bias voltages are 
shown in Fig. \ref{Fig3}. 
Top panels represent simulated STM topography of 3 $\times$ 3 nm$^2$ of the 
same area of the Si(335)-Au/Pb surface for sample bias $U=-1$ V (a) and 
$U=-0.1$ V (b). For comparison, corresponding images of the same area of clean 
Si(335)-Au surface are shown in the bottom panels. As it was discussed 
previously (see Sec. \ref{clean}), the STM topography of clean Si(335)-Au 
surface features monoatomic chains, which are associated with the step edge Si 
atoms (see Fig. \ref{Fig3}c) and d)). The Pb atoms deposited on this surface 
form monoatomic chains, located in the middle of terraces, i.e. between the 
chains of clean Si(335)-Au surface. This is evident, if one compares panels a) 
and c) or b) and d) of Fig. \ref{Fig3}.  

\begin{figure}[h]
 \resizebox{0.9\linewidth}{!}{
  \includegraphics{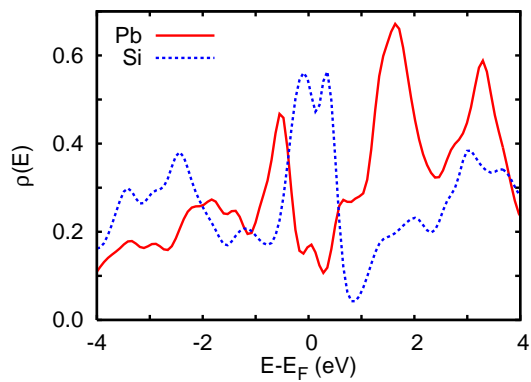}
} 
 \caption{\label{Fig4} (Color online) Projected density of states of the Pb
          (solid line) and the step edge Si atom (dashed line).
 }
\end{figure}

As one can read off from Fig. \ref{Fig3}, the Pb atoms are more pronounced than 
the step edge Si atoms, especially at high sample bias. Such a behavior can be 
explained by the combination of structural end electronic effects. The Pb chain 
sticks out (by 2 \AA $\;$) above the surface, and this mainly contribute to the 
discussed effect. On the other hand, the electronic properties also play 
significant role. The calculated projected density of states (PDOS) of the Pb 
atoms and the step edge Si atoms is shown in Fig. \ref{Fig4}.
The Pb PDOS features larger values at energies $- 0.7$ eV and $+ 1.5$ eV (see 
solid line in Fig.\ref{Fig4}), while PDOS of the step edge Si atoms is larger 
near the Fermi energy (dashed line). In first approximation, the STM current is 
proportional to integrated density of states between the Fermi energy and 
applied bias voltage (eU). This contributes to the fact that the Pb chain is
more pronounced at higher sample bias. At very low sample bias ($U = -0.1$) V, 
the step edge Si atoms have larger values of PDOS, and as a result both chains 
have comparable topography amplitude (see panel b) of Fig. \ref{Fig3}). 
However, the Pb chain is still more pronounced than the Si one due to the fact 
that it sticks out above the surface. 

The STM topography along the Pb chain shows very small changes of the amplitude 
at low voltages (see panel b) of Fig. \ref{Fig3}), indicating strong 
overlapping of the 6p states. This is also reflected in the band structure, 
where 6p band of lead, crossing the Fermi energy, is very dispersive in the 
direction along the chain (see Fig. \ref{Fig4}). On the other hand, there is no 
such an effect in the STM topography of the step edge Si atoms due to more 
localized character of 3p states of silicon. As a result the step edge Si band 
is rather flat, i.e. is less dispersive than 6p band of lead (see Fig. 
\ref{Fig4}). All this shows that the Pb chain has more metallic character than 
the step edge Si chain, which makes the lead a good candidate to look for 
exotic phenomena characteristic for the systems of reduced dimensionality.


\section{\label{band} Band structure}

\begin{figure}[h]
 \resizebox{\linewidth}{!}{
  \includegraphics{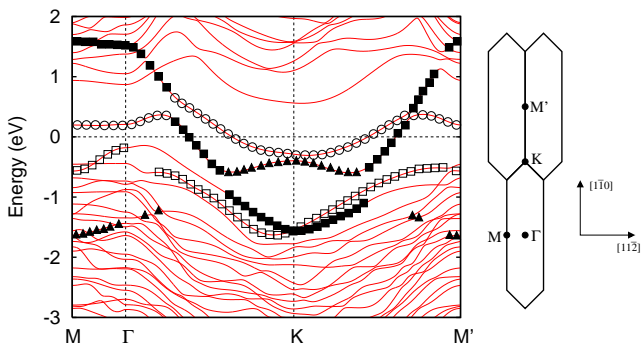}
} 
 \caption{\label{Fig5} (Color online) Calculated band structure along high 
          symmetry lines in two-dimensional Brillouin zone (shown on the 
	  right). The $[1 \bar{1} 0]$ direction is parallel to the steps, while 
	  $[1 1 \bar{2}]$ - perpendicular to them. The bands associated with Pb 
	  chain is marked with filled triangles, the step edge Si atom band - 
	  open circles, while the bands originating from the Au chain 
	  hybridizing with Si$_4$ and Si$_5$ atoms are marked with squares. The 
	  Fermi level is taken as the zero of energy.
 }
\end{figure}

The calculated band structure for present structural model, along the high
symmetry lines of two-dimensional Brillouin zone (BZ), is shown in Fig. 
\ref{Fig5}.
The direction in the 2D Brillouin zone, defined by the points $\Gamma$, K and 
M', is parallel to the steps of the Si(335)-Au surface, while the $\Gamma$-M is 
perpendicular to them (see right panel of Fig. \ref{Fig5}). A few important 
surface bands are marked with symbols in Fig. \ref{Fig5}. The band marked with 
open circles pinning the Fermi energy, originates from unsaturated bonds of the 
Si atoms at the step edge. The band 0.5-1.5 eV below the Fermi energy (open 
squares), as well as more dispersive band crossing the E$_F$ (filled squares) 
are associated with the Au chain. To be more precise, those bands originate 
from the hybridization of the Au chain with neighboring Si$_4$ and Si$_5$ atoms 
(see Fig. \ref{Fig2}). All the bands discussed above have also been identified 
in the clean Si(335)-Au surface \cite{MK_2} and similar bands have been found 
in the Si(557)-Au reconstruction \cite{Sanchez_1,Sanchez_2}. In fact, the band 
marked with filled squares has also Pb character. This band reflects the 
hybridization of gold with Pb and Si$_5$ atoms. Finally, the band shown as 
filled triangles is associated with 6p states of lead. All the above bands do 
not have (or have very weak) dispersion in the direction perpendicular to the 
steps ($\Gamma$-M), indicating one-dimensional character of the structure.

Since Pb features a strong spin-orbit (SO) interaction \cite{Dil,Bihlmayer}, 
and the low-dimensionality of the system can increase it \cite{Agrawal}, it is 
worthwhile to comment on this effect in the Si(335)-Au/Pb system. Although the 
spin-orbit interaction was not included in present calculations, one can draw 
some conclusions appealing to the Si(557)-Au reconstruction. The measured band 
structure of Si(557)-Au surface shows two Au induced proximal bands crossing 
the E$_F$ near the K point (of $ 2 \times 1$) zone with $\sim$ 300 meV 
splitting \cite{Losio,Altmann,Ahn,Kisiel}. An explanation of the splitting was 
given in terms of SO interaction \cite{Sanchez_3}. Since both Si(335)-Au and 
Si(557)-Au surfaces belong to the same family of vicinal surfaces (Si(335)-Au 
reconstruction may be considered as a simple truncation of Si(557)-Au surface), 
one can expect SO interaction to play similar role here. In fact, Crain 
{\it et al.} \cite{Crain} observed small splitting of the band crossing the 
E$_F$. This was interpreted as two different bands, one coming from Au-Si 
hybridization and the other one from the step edge Si atoms \cite{MK_2}. It is 
also possible that the splitting has its origin in the SO interaction, similar 
like in Si(557)-Au surface. On the other hand, if Pb is deposited onto 
Si(335)-Au surface, one would expect SO effect to be more pronounced. However, 
the ARPES data \cite{Kisiel_2} shows a single metallic band and thus no 
evidence of SO splitting. So one can conclude that if SO effect is really 
present in this system, certainly the splitting of the band is smaller than the 
energy resolution of the ARPES apparatus, which is 50 meV in this case. This is 
rather an intriguing result, however it can be true, as recent spin-polarized 
photoemission measurements report values of SO splitting as small as 15 meV in 
Pb deposited on Si(111) surface \cite{Dil_2}. 

\begin{figure}[h]
 \resizebox{0.9\linewidth}{!}{
  \includegraphics{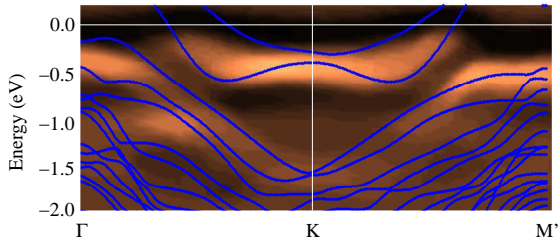}
} 
 \caption{\label{Fig6} (Color online) Comparison of the measured ARPES 
          intensity along $\Gamma$-K-M' line of the 2D Brillouin zone 
	  (Ref. \cite{Kisiel_2}) and the calculated band structure. The high 
	  intensity is shown light.
 }
\end{figure}

\begin{figure}[h]
 \resizebox{0.9\linewidth}{!}{
  \includegraphics{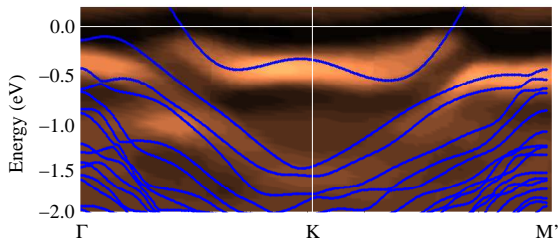}
} 
 \caption{\label{Fig7} (Color online) Comparison of the measured ARPES 
          intensity (Ref. \cite{Kisiel_2}) and the calculated band structure 
	  for the model with saturated step edge Si bonds. 
 }
\end{figure}

A comparison of the calculated band structure with the photoemission spectra of
Ref. \cite{Kisiel_2} is shown in Fig. \ref{Fig6}.
As one can see, the present DFT calculations agree very well with the
experimental data. In particular, the band near the Fermi energy, which has 
high photoemission intensity, is the 6p band of lead. One can also identify 
other bands, marked with symbols in Fig. \ref{Fig5}, except the band coming 
from the step edge Si atoms. The reason that this band is not observed in ARPES 
can be twofold. First, the band is to close to the Fermi energy which makes it 
very difficult to verify experimentally. Second, there may be some mechanism 
which leads to the saturation of the Si bonds, accounting for various 
imperfections or impurities. It is more likely that the band associated with 
the unsaturated step edge bonds (open circles in Fig. \ref{Fig5}) is eliminated
in the real surface by some reconstruction and passivation. The main effects of
the reconstruction can be achieved by saturating the dangling bonds with
hydrogen \cite{Sanchez_1}. This process removes extra band from the gap region 
(open circles in Fig. \ref{Fig5}), and is energetically very favorable, as the 
energy gain is $\sim 1.30$ eV with respect to the unsaturated surface and H$_2$ 
molecule. The comparison of the calculated band structure with saturated Si$_1$
bonds and the ARPES spectra of Ref. \cite{Kisiel_2} is shown in Fig. 
\ref{Fig7}.
Clearly, the band structure is in better agreement with the ARPES experiment
now. This suggests that the step edge Si bonds are really saturated by some 
surface reconstruction, however this cannot be accounted for without 
multiplying the unit cell of the Si(335)-Au surface.


\section{\label{buckling} Step-edge buckling}


The Si(557)-Au reconstruction is known to undergo a buckling of the step edge at
low temperatures \cite{Sanchez_2,Sanchez_3,Riikonen}. The step edge Si atoms 
alternate between up and down positions, with distortion in z direction 
$\Delta z = 0.65$ \AA. In general, all the step edge atoms occupy equivalent 
positions and feature half-filled dangling bonds, which give rise to a flat 
band near the Fermi energy. This is referred as high-temperature phase. However 
this situation is energetically unfavorable at low temperature and the system 
tends to lower its energy by the buckling of the step edge. The energy gain is 
130.9 meV. The dangling bonds of the up-edge Si atoms became fully occupied, 
while those of the down-edge are empty. As a result the band structure features 
the fully occupied and empty step-edge electron bands. The buckling of the step 
edge influences the properties of the band structure near the Fermi energy 
through changes in the Au-Si-Au bond angles. Those changes are responsible for 
the band gap that opens in dispersive Au-Si band crossing the E$_F$, and thus 
for the metal-insulator transition at low temperature \cite{Riikonen}.

\begin{figure}[h]
 \resizebox{0.9\linewidth}{!}{
  \includegraphics{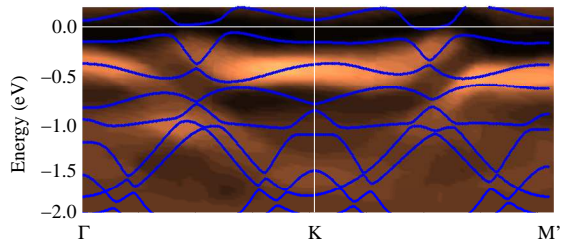}
} 
 \caption{\label{Fig8} (Color online) Comparison of the measured ARPES 
          intensity (Ref. \cite{Kisiel_2}) and the calculated surface band 
	  structure for the model with the buckling of the step edge. 
 }
\end{figure}

Doubling the unit cell of Si(335)-Au/Pb system in direction $[1 \bar{1} 0]$ one 
can account for similar effect associated with the buckling of the step edge. 
In the present system the distortion $\Delta z$ is slightly smaller than in
Si(557)-Au system, and equal to 0.57 \AA. This smaller value of the $\Delta z$
can be related to the fact that Si(557)-Au supercell cell consists of two 
non-equivalent $\times 1$ unit cells with extra Si adatom in one of them, which
in turn can increase $\Delta z$. The $\times 2$ Si(335)-Au/Pb supercell is 
constructed from two equivalent $\times 1$ cells. The buckling of the step edge 
lowers the total energy of the system by 117.5 meV, to be compared with 130.9 
meV in the case of Si(557)-Au surface, and leads to a splitting of the step 
edge Si band, originally pinning the Fermi energy. As a result one observes two 
bands, a fully occupied flat band (with a bandwidth $W \approx 0.09$ eV) at 
energy 0.29 eV below E$_F$, and empty more dispersive band ($W \approx 0.38$ 
eV) at energy 0.25 eV above the E$_F$. Those bands are associated with up-edge 
and down-edge Si atoms, respectively. Similar bands have also been observed in 
the Si(557)-Au surface \cite{Riikonen}. The step edge buckling is also 
reflected in different bond angles between Au and Si$_5$ atoms (see Fig. 
\ref{Fig2}). In situation when all the step edge Si atoms occupy equivalent 
positions ($\Delta z = 0$), these angles are equal 101.64$^{\circ}$, to be
compared with 111.6$^{\circ}$ and 103.7$^{\circ}$ for Si(557)-Au surface. 
Again, different values of the Si(557)-Au angles come from non-equivalent
$\times 1$ unit cells, and the buckling can change them by $\pm 10^{\circ}$
\cite{Riikonen}. On the other hand, the buckling in the present system changes 
the Si$_5$-Au-Si$_5$ angles to 101.75$^{\circ}$ (up-edge) and 101.29$^{\circ}$
(down-edge $\times 1$ cell). Moreover, the angles between Au and Pb atoms are 
also slightly changed, and equal 81.78$^{\circ}$ (81.54$^{\circ}$) in up-edge 
(down-edge) $\times 1$ cells, to be compared with equilibrium value 
81.97$^{\circ}$. All this is reflected in the band structure, where very small 
energy gap develops in dispersive Au-Pb-Si band (filled squares in Fig. 
\ref{Fig5}), when the step-edge buckling takes place. However, this small 
energy gap is slightly below the Fermi level. Similar effect, although with
slightly larger energy gap, has been found in Si(557)-Au surface 
\cite{Riikonen}, and has been assigned to the shortcomings of the LDA 
functionals, which fail to describe excitation spectra. So one can expect that 
better description of the exchange and correlation effects (like GW 
approximation) will move the empty states to higher energies, thus driving the 
system into insulating state \cite{Rohlfing,Faleev}. However, it is also
possible, that present LDA results are correct, and the system stays in 
metallic phase. The step-edge buckling may not be able to open a true gap in 
Au-Pb-Si band, as it gives much smaller values of Au-Si angles than in case of
the Si(557)-Au surface. This last scenario seems to be supported by 
experimental data of Ref. \cite{Kisiel_2}, shown in Fig. \ref{Fig8} together
with calculated surface band structure for the step-edge buckled Si(335)-Au/Pb
reconstruction.
As one can see the agreement between measured and calculated band structure is 
very good. In particular there is no band pinned at the Fermi energy. 

To calculate the surface band structure shown in Fig. \ref{Fig8}, following 
procedure has been applied. Since the bulk periodicity in direction 
$[1 \bar{1} 0]$ is $1 \times a_{[1 \bar{1} 0]}$, and calculations have been 
performed in $\times 2$ unit cell in this direction, the resulting band 
structure features twice as many bands as in $\times 1$ unit cell, folded back 
into first BZ of $\times 2$ structure. This of course is not observed in 
experiment, as the ARPES is surface sensitive technique. To get rid of these 
bulk electron bands, only the surface atoms have been taken into account in
calculations. The positions of these atoms have been obtained in full slab
calculations, as described in Sec. \ref{details}. To mimic rest of the Si 
layers, the surface Si atoms were passivated by hydrogen. The H atoms were
placed in a way that the bond lengths and bond angles between surface and next 
from the surface Si layers were preserved. In this way calculated band 
structure is shown in Fig. \ref{Fig8}. Note that differences between in this 
way calculated surface band structure and that obtained from full slab
calculations are negligible.


\section{\label{conclusions} Conclusions}


In conclusion, the structural and the electronic properties of the monoatomic 
Pb chains on Si(335)-Au surface have been studied within the density functional
theory. The obtained structural model features Pb chain on-top of the surface 
in the middle of terrace. As a result, two monoatomic chains are observed on
single terrace, one made of lead, and the other one associated with the step
edge Si atoms. Both chains are visible in STM experimental data. The STM 
topography is very well reproduced within the present model. The calculated 
band structure shows clear one-dimensional character of the structure and 
remains in almost perfect agreement with the ARPES data. Finally, the buckling 
of the step edge Si atoms has been found, and unlike for the Si(557)-Au 
surface, does not drive the system into insulating phase.


\begin{acknowledgments}
I would like to thank Prof. M. Ja\l ochowski for valuable discussions. This 
work has been supported by the Polish Ministry of Education and Science under 
Grant No. N202 081 31/0372.
\end{acknowledgments}



\end{document}